\documentclass[jocse,table]{jocseart}
\usepackage[table]{xcolor}

\usepackage{booktabs} % For formal tables
\usepackage{graphicx}
\usepackage{amsmath}

\usepackage{subcaption}
\setcopyright{jocsecopyright}

\jocseDOI{10.22369/issn.2153-4136/x/x/x }

\begin{document}

\title[Captive Portal Technology: A Modular, Hands-On Approach to Infrastructure]{Integrating Captive Portal Technology into Computer Science Education: A Modular, Hands-On Approach to Infrastructure}

\author{Lianting Wang}
\email{lianting.wang@alumni.utoronto.ca,lianting.wang@outlook.com}
%\email{lianting.wang@mail.utoronto.ca}
\affiliation{%
  \institution{Department of Computer and Mathematical Sciences, University of Toronto Scarborough}
  \city{Toronto}
  \state{ON}
  \country{Canada}
}

\author{Marcelo Ponce}
\email{m.ponce@utoronto.ca}
\affiliation{%
  \institution{Department of Computer and Mathematical Sciences, University of Toronto Scarborough}
  \city{Toronto}
  \state{ON}
  \country{Canada}
}

% Abstract 
\begin{abstract}
In this paper, we present an educational project aimed to introduce students to the
technology behind \textit{Captive Portals} infrastructures.
For doing this, we developed a series of modules to emphasize each of the
different aspects and features of this technology.
The project is based on an open source implementation which is widely used
in many computer network courses, making it well-suited and very appealing
for instructors and practitioners in this field.

\end{abstract}

%%%%%%%%%

% Keywords



%%%%%%%%%

% Generate the title
\maketitle

% Input the body of the paper by providing the file name. For example,
% if the file name is body.tex, then input{body}
%Please use this simple set up -- makefiles or other structures will cause delays in publication.

\section{Introduction}\label{introduction}

A \textit{Captive Portal} is a form of network connectivity where access to the
Internet is restricted until certain requirements are satisfied.
It presents the access-point provider's requirements through an interface,
typically a web-page type, such as reading advertisements, accepting acceptable
usage policies, or providing some form of credentials \cite{larose2020rfc}.

Nowadays, Captive Portals are ubiquitous in our lives. Whether we are in
airports, hotels, or restaurants, Captive Portals have always provided an
easy way to authenticate our identity and access the Internet. Although
the time spent interacting with Captive Portal interfaces is short and in some
cases considered a formality, the technology encompasses a rich set of
network protocols and authentication processes.
HPC and distributed compute systems can use this approach to validate users and access to resources, e.g. gateways to specialized portals, disclose systems' policies, etc.
Automated pop-ups from a
mobile phone or computer device further illustrate the versatility and
complexity of this technology. These features will stimulate people's
curiosity and encourage them to further explore the field of computer
network security and infrastructure \cite{10.1007/978-3-030-31500-9_6}.

The pervasive nature of Captive Portals in everyday scenarios presents a
unique opportunity for computer science educational exploration,
especially in the field of computer networks. Students are already familiar with the
user-oriented aspects of the technology and are standing on the
threshold of deeper exploratory learning. By demystifying the core
principles of Captive Portals, we can transform students' casual
encounters with this technology into profound learning experiences. This
educational journey promises not only to help students gain a deeper
understanding of network communications and its related protocols, but
also to foster critical thinking about the security measures inherent in
seemingly mundane digital interactions.

The teaching case model proposed in this paper is an innovative
educational tool designed to explore Captive Portal technology in depth
through hands-on practice, thereby promoting students' comprehensive
understanding of basic network technologies. By simulating the operation
of a typical Captive Portal, students are able to intuitively learn and understand
the core technical principles in networking, including the reasons for
the use of media access control (MAC) addresses, the role of address resolution protocol (ARP) requests,
the working mechanism of switches, the processing of domain name system (DNS) requests,
the methods of applying web servers, and the key differences between HTTP and HTTPS.
This type of learning is not limited to academic exploration; it also
helps students relate their theoretical knowledge to the construction
and operation of networks in their daily lives, leading to a more
intuitive understanding of how networks work.

This model emphasizes the close integration of technology and life,
encouraging students to become more engaged in their studies by
increasing their interest, while also developing their practical
application skills \cite{10.1145/3328778.3366891}.
Through this model, students are able to not
only grasp the technical fundamentals of the Captive Portal system, but
also critically assess its security and further understand the
importance of network security and infrastructure. This pedagogical
approach seeks to bridge theoretical knowledge with its practical
application, enabling students to apply what they have learned to solve
real-life network problems they may encounter, thereby developing their
comprehensive understanding and application skills.

\begin{comment}
In conclusion, the development and implementation of a Captive Portal
model for educational purposes embodies a strategic initiative to enrich
computing education. By integrating comprehensive technical content
related to computer networks, this model not only illuminates the
operational essence of Captive Portals but also paves the way for
innovative educational practices in the field of computing and
information technology.
\end{comment}

%%%%%%%%%%%%%%%%%%%%%%%%%%%%%%%%%%%%%%%%%%%%%%%%%%%%%%%%%%%%%%%%%%%%%

\section{Design Principles}\label{design-principles}

\subsection{Introduction to Design Principles}\label{introduction-to-design-principles}

The Captive Portal instructional project is underpinned by a set of core
design principles aimed at maximizing the educational value for
students. These principles are crafted to ensure a blend of practical
learning experiences with a strong theoretical foundation. By engaging
students in hands-on activities involving Software-Defined Networking
(SDN) and Captive Portal technologies, the project bridges the gap
between conceptual knowledge and real-world applications. The
overarching goal is to foster a deep understanding of network
communication principles, system architecture, and the challenges of
implementing network-based applications.

\subsection{Extensive Coding Instructions}\label{extensive-coding-instructions}

Central to our approach is the provision of detailed coding instructions
and guidelines. Before embarking on code development, students are
required to engage with comprehensive documentation that outlines the
system's architecture, design considerations, and specific module
requirements. This preparatory step is designed to shift the focus from
minutiae to the broader system perspective, encouraging students to
appreciate the interconnectedness of system components. Supplementary
resources, including helper functions and baseline code, are provided to
streamline the implementation process, enabling students to concentrate
on solving more complex problems.

\subsection{Modular Design}\label{modular-design}

Acknowledging the diverse focus areas and difficulty levels required
across different courses, the Captive Portal project is inherently
modular. This flexibility allows instructors to tailor the project to
fit educational objectives, selecting modules that align with specific
aspects of networking, such as SDN or DNS. Moreover, the modularity
supports scalability in complexity, making it suitable for a wide range
of student expertise. Such a design not only enhances the learning
experience but also empowers educators to customize the curriculum to
meet students' needs more effectively.

\subsection{Critical Thinking and Problem-Solving}\label{critical-thinking-and-problem-solving}

The tiered difficulty levels within the modular design introduce a
unique opportunity for critical evaluation. Students are challenged to
assess the adequacy of the current module design, promoting critical
thinking. This evaluative process deepens their understanding of the
system, encouraging a thoughtful consideration of design efficiency,
scalability, and performance. Through this, students develop a keen eye
for optimizing solutions, a skill critical in the fields of HPC and technology
development \cite{heijltjes2015unraveling}.

\subsection{Stable Development Environment}\label{stable-development-environment}

Recognizing the complexity of setting up a development environment for
such a diverse project, we provide a pre-configured virtual environment.
This environment includes all necessary dependencies and tools, ensuring
uniformity across all modules. By eliminating the setup barriers,
students can immediately focus on the learning objectives and practical
experience, rather than grappling with configuration challenges.

\subsection{Test-Driven Development}\label{test-driven-development}

Embracing best practices in software development, the project emphasizes
test-driven development (TDD). For each module, comprehensive test
suites are provided, guiding students through the development process.
This approach not only facilitates immediate feedback on the correctness
of implementations but also instills a disciplined coding methodology \cite{10.5555/1040231.1040261}.
By integrating TDD, students learn to prioritize functionality
and reliability from the outset, a practice that significantly benefits
their future professional endeavors.

%%%%%%%%%%%%%%%%%%%%%%%%%%%%%%%%%%%%%%%%%%%%%%%%%%%%%%%%%%%%%%%%%%%%%

\section{Implementation}\label{implementation}

For our project we decided to use an open-source tool named \emph{mininet}
--\url{https://mininet.org/}-- \cite{7311238,XIANG2020219}, which has been
widely used and adapted in many computer networks courses due to its
versatility, flexibility and open approach.
Employing \emph{mininet} as a \textit{software defined network} (SDN) solution,
we developed our main infrastructure for the teaching modules and educational
projects.

\begin{figure}
%\centering
	\includegraphics[width=\columnwidth]{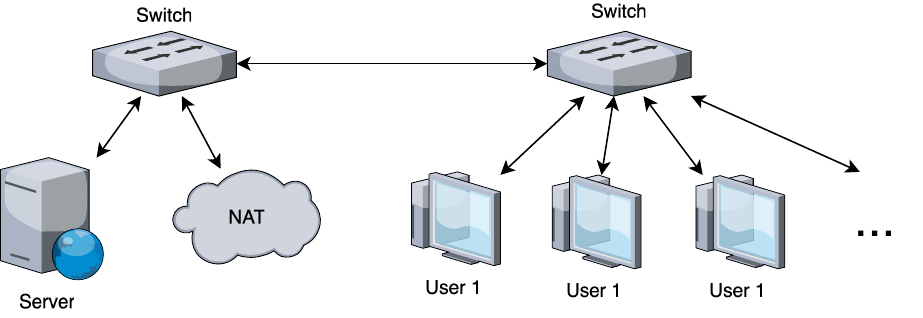}
	\caption{Basic structure for a typical Captive Portal configuration.}
	\label{fig:min_CaptPort}
\end{figure}

Fig.~\ref{fig:min_CaptPort} depicts a typical infrastructure for a Captive Portal
deployment, where users and servers are connected through two Switches.

Using SDN technology, we can manage connections at the link layer, directing
users with unique MAC addresses either to the network address translation (NAT)
or to the Captive Portal server. However, since the network layer focuses solely
on delivering data to the correct IP address without regard to the
connection's status, we employ specific techniques to reroute user
requests to the appropriate server.

\begin{figure*}
\centering
	\includegraphics[width=.7\textwidth]{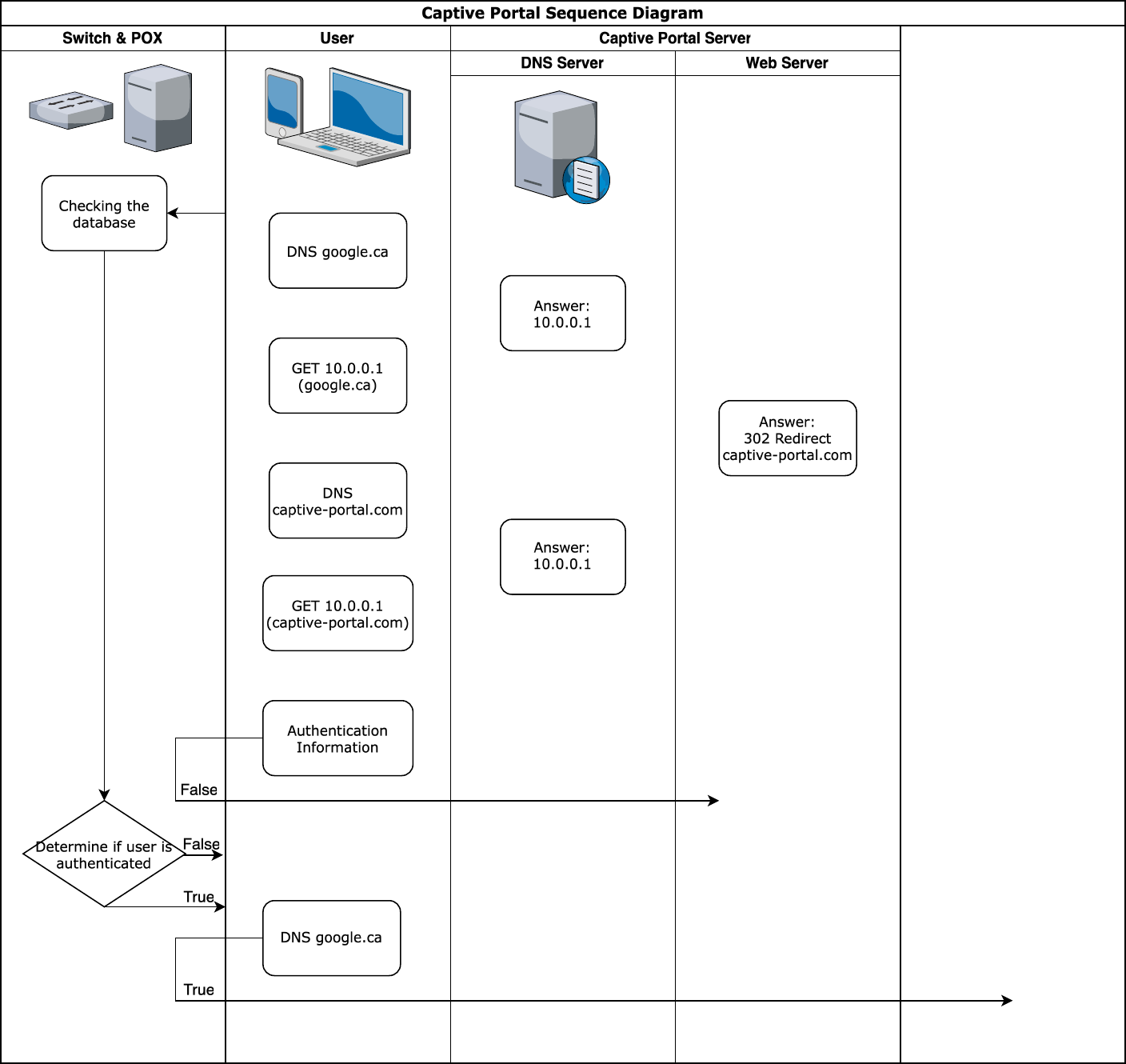}
	\caption{Sequence Diagram of Captive Portal with DNS Spoofing Technology}
	\label{fig:CP_wDNSspoofing}
\end{figure*}

One possible method is the so-called \textit{DNS Spoofing} -- see Fig.~\ref{fig:CP_wDNSspoofing}.
In this case when a user tries to access an external website, their device
sends out a DNS request to find the website's IP address.
Normally, this would lead them to the intended website, but we intervene by using a special DNS server. This
server directs all domain name requests to the Captive Portal server,
ensuring the user encounters the authentication page regardless of their
intended destination.

Another approach we use is \textit{IP Forgery}. With this technique, when a
user`s device sends a DNS request, we provide the correct IP address.
However, when the user attempts to access the website, we intercept the
web request and respond with an HTTP redirect message that redirects the
user to the Captive Portal server's domain name. The user`s second
connection will connect to the Captive Portal server.

\begin{figure*}
\centering
	\includegraphics[width=.685\textwidth]{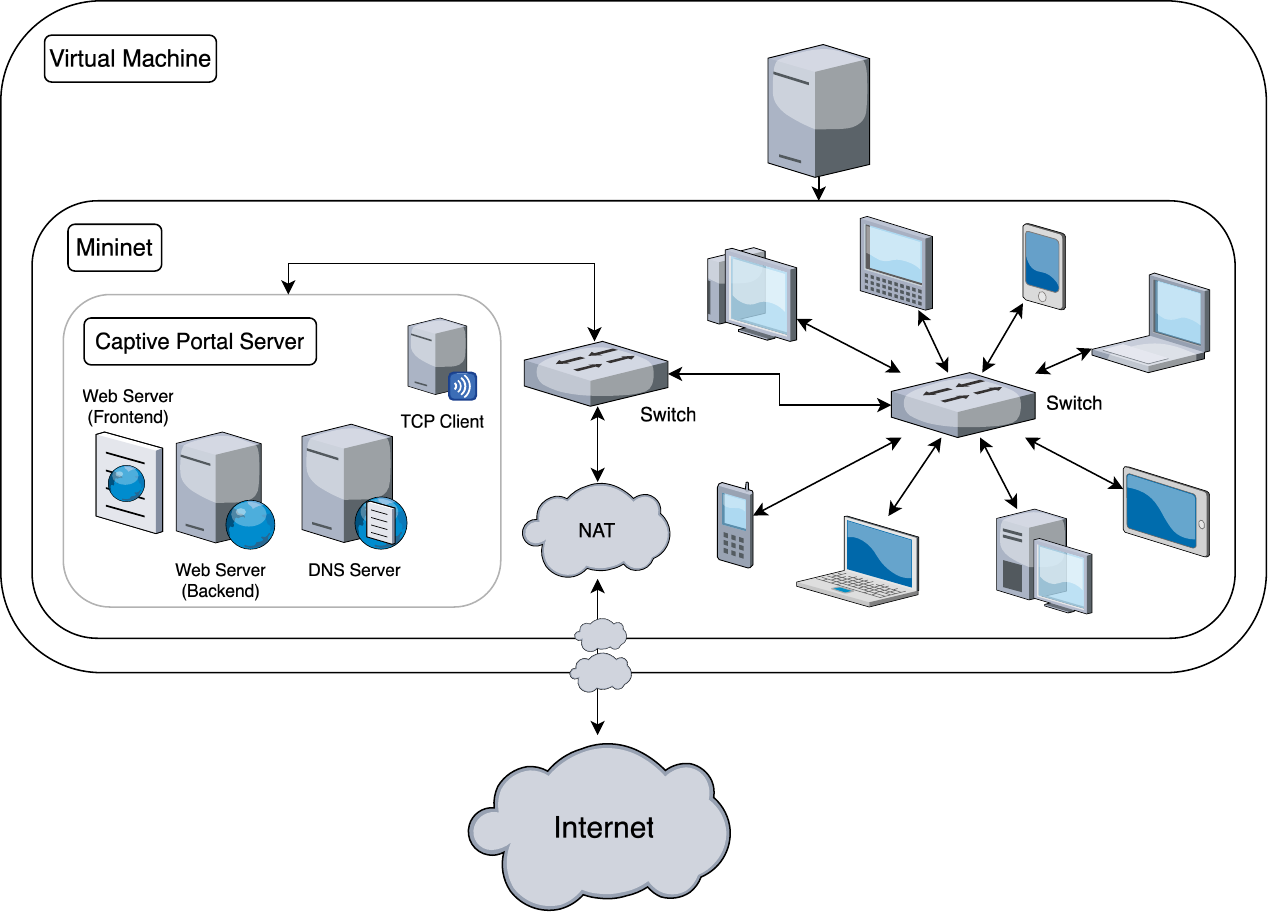}
	\caption{Framework of Captive Portal project.}
	\label{fig:CP_framework}
\end{figure*}

\begin{figure}
	\includegraphics[width=.6\columnwidth]{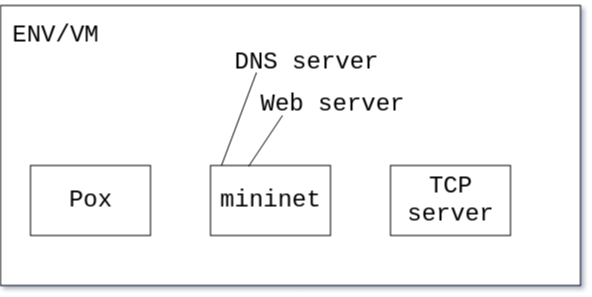}
	\caption{Simplified view of the main components of the Captive Portal setup as
	described in Fig.~\ref{fig:CP_framework}: POX controler, mininet SDN emulator, and TCP server.}
	\label{fig:CP_diagram}
\end{figure}

Regardless of the specific methodology used, the framework of the entire
project is fixed, as shown in Figs.~\ref{fig:CP_framework} and ~\ref{fig:CP_diagram}.

%%%%%%%%%%%%%%%%%%%%%%%%%%%%%%%%%%%%%%%%%%%%%%%%%%%%%%%%%%%%%%%%

\section{Modules}\label{modules}

In this section we describe the main modules/projects we have originally designed with
the goal of guiding interested readers through the different instances of 
the Captive Portal technology.

The details of the implementations, along with specific examples, demos and
any software and configuration details are available in our public repository,
	\url{https://github.com/Lianting-Wang/Captive-Portal-Education}.

Further details about these are also available in Appendix~\ref{ap:modules_details}.

%%%%
\begin{itemize}
	\item \textbf{Module 1 - TCP Sever and Client}:
		%\label{module-1---tcp-sever-and-client}
		%\\
This module lays the foundational understanding of TCP connections,
vital for implementing captive portal projects. It aims to solidify
students' network programming skills and introduce them to Python coding
for network applications.

%%%%

	\item \textbf{Module 2 - SDN Switch}:
		%\label{module-2---sdn-switch}
		%\\
Focuses on understanding Software-Defined Networking (SDN) controllers
and the manipulation of network traffic. Through hands-on practice with
the POX SDN Controller, students will learn how to program switches and
process network packets.

%%%%%

	\item \textbf{Module 3 - DNS Server}:
		%\label{module-3---dns-server}
		%\\
Introduces the functionality and implementation of DNS servers within
the context of captive portal technology. This module covers different
approaches to handling DNS requests, emphasizing hands-on experience
with DNS manipulation.

%%%%%

	\item \textbf{Module 4 - Web Server}:
		%\label{module-4---web-server}
		%\\
Explores the deployment and configuration of a local web server to
interact directly with network requests, crucial for the operation of a
captive portal.

%%%%%

	\item \textbf{Module 5 - Mininet Script}:
		%\label{module-5---mininet-script}
		%\\
Delves into network simulation using Mininet, teaching students to
construct and modify network topologies to route user requests through
specific servers, mimicking a captive portal environment.

%%%%%%

\end{itemize}

%%%%%%%%%%%%%%%%%%%%%%%%%%%%%%%%%%%%%%%%%%%%%%%%%%%%%%%%%%%%%%%%

\section{Code Provisioning Tools}\label{code-provisioning-tools}

To facilitate a streamlined educational experience in Captive Portal
development, we have implemented a comprehensive suite of code
provisioning tools, designed to integrate seamlessly into academic
settings. This suite includes a meticulously crafted flowchart
--see Fig.~\ref{fig:CP_flowchart}-- and an
accompanying website, which together serve as a central hub for
instructional resources and starter code packages tailored to various
teaching objectives.
These tools, in addition to other elements, like demos and tutorials
can be found in our publicly available repository:

	\url{https://github.com/Lianting-Wang/Captive-Portal-Education}.

In addition to that, we have developed an automated script to assist users with
the setup process, as well as, an automated test script, which also verifies
whether the environment is configured correctly.

\subsection{Flowchart Design}\label{flowchart-design}

\begin{figure*}
	\includegraphics[width=\textwidth]{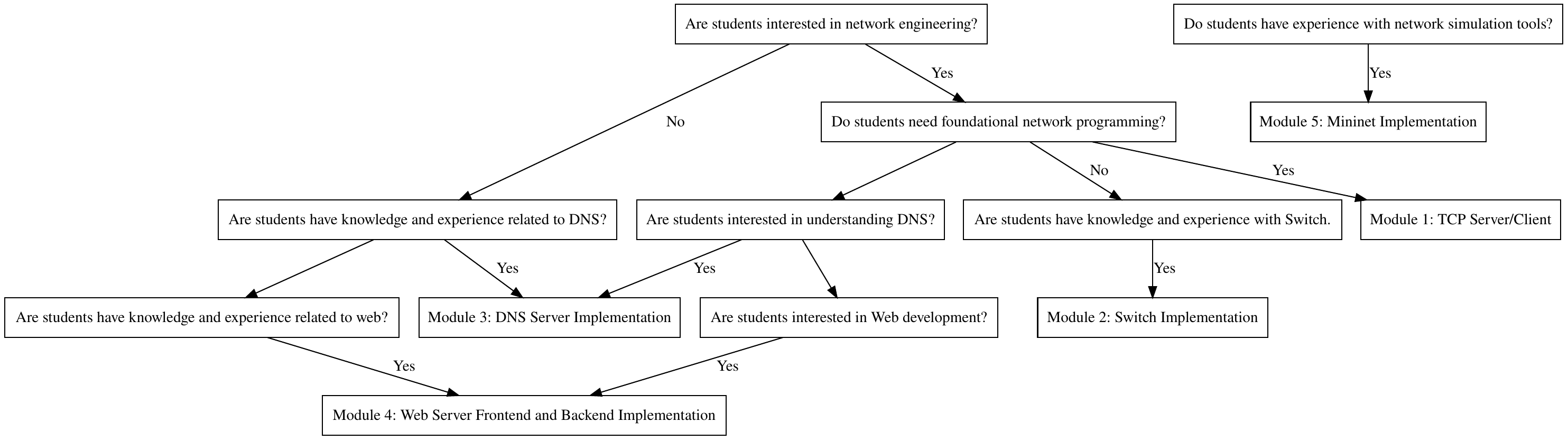}
	\caption{Captive Portal education modules \textit{recomendation system} -- flowchart used to determine what modules are recommended to use based on the desired educational targets and goals.
	Also available as an interactive tool in the web-interface from the Captive Portal educational repository.}
	\label{fig:CP_flowchart}
\end{figure*}

The flowchart, Fig.~\ref{fig:CP_flowchart}, is engineered to guide educators through a structured
selection process, enabling them to identify and choose the most
appropriate instructional materials and starting code bases that align
with their specific teaching goals and the learning outcomes desired for
their students. This visual tool simplifies the complexity of course
design by categorizing materials based on topics, difficulty levels, and
the progression of concepts essential to Captive Portal development.

\subsection{Website Interface}\label{website-interface}

Complementing the flowchart, the website offers an intuitive,
user-friendly platform where educators can effortlessly download
pre-packaged instructional materials and starting code. This eliminates
the traditionally time-consuming process of manual compilation and
customization of teaching resources. Educators can make selections based
on the guidance provided by the flowchart, with the website offering
detailed descriptions, download links, and additional resources for each
package.

\subsection{Key Features}\label{key-features}

\begin{itemize}
%\tightlist
\item
  Customization: Educators can customize packages to suit the unique
  needs of their curriculum, allowing for flexibility in teaching
  advanced topics or focusing on foundational skills.
\item
  Accessibility: The website ensures easy access to up-to-date
  materials, fostering an adaptive learning environment that can quickly
  respond to the evolving landscape of Captive Portal technologies.
\end{itemize}

%%%%%%%%%%%%%%%%%%%%%%%%%%%%%%%%%%%%%%%%%%%%%%%%%%%%%%%%%%%%%%%%%%%%%

\section{Discussion}\label{discussion}

The transition to environments supportive of Apple's M-series chips has
unveiled challenges, notably in virtualization capabilities crucial for
educational settings. Our ongoing efforts to integrate Docker as a
foundational component for our instructional environment have
encountered obstacles, particularly with the integration of the X Window
System. Despite these hurdles, we remain optimistic about resolving
these issues, ensuring a seamless experience across diverse hardware
platforms.

An alternative to be considered for Apple silicon, would be alternative
emulation environments like the one offered by EMU (\url{https://getutm.app/}) or UTM (\url{https://www.qemu.org/}).
In these cases, it would require porting the whole infrastructure associated
to the SDN layer and additional components of the Captive Portal implementation.

In the past, we have also students attempting a fresh installation and compilation
of mininet on native Mac OS with relative success although this it not
complexly clear is a solution that implies a substantial efforts commitment as well as
is not clear whether it will be sustained in time.

We believe this is important information to share from educators perspective,
as it has proven to be an additional annoyance for some students to deal with.
The palliative solution we offer some of our students is to use Linux based lab. machines,
which can be accessed in-situ as well as remotely -- for in which cases, we will
teaching technologies such as remote connectivity via ssh, X-forwarding, VNC and even
VPN due to universities policies and cyber-security best practices \cite{jocse-14-2-3}.

In addition to platform compatibility concerns, our evaluation of
Mininet --sometimes also considered as a lightweight network simulation tool-- has revealed
limitations in its capacity to simulate hardware intricacies accurately.
This shortfall has prompted a quest for more sophisticated alternatives
capable of delivering a purer simulation experience. Our aim is to
enrich the educational experience by offering students hands-on exposure
to configuring a variety of network protocols from the ground up. This
endeavor seeks to bridge theoretical knowledge with practical
application, enhancing the comprehensiveness of our instructional
approach.

%%%%%%%%%%%%%%%%%%%%%%%%%%%%%%%%%%%%%%%%%%%%%%%%%%%%%%%%%%%%%%%%%%%%%

%%%%%%%%%%%%%%%%%%%%%%%%%%%%%%%%%%%%%%%%%%%%%%%%%%%%%%%%%%%%%%%%%%%%%

\section{Conclusions}

In this work we present a module-based project aiming to familiarize students
with Captive Portal technologies. We do so, by using a series of small problems
employing the mininet SDN tool.

Our project is open source and hosted in publicly available repository, emphasizing
relevant aspects of its open source approach, such as reproducibility, modularity
and sustainability.
Furthermore, because of its openness, it is possible to continue to grow this project,
by adding more modules, accepting contributions from the community, etc.

In conclusion, the development and implementation of a Captive Portal model for
educational purposes embodies a concrete contribution to enrich computing
education. By integrating comprehensive technical content related to computer
networks, this model not only illuminates the operational essence of Captive
Portals but also paves the way for additional educational practices in the
field of computing and information technology.

%%%%%%%%%%%%%%%%%%%%%%%%%%%%%%%%%%%%%%%%%%%%%%%%%%%%%%%%%%%%%%%%%%%%%

%%%%%%%%%%%%%%%%%%%%%%%%%%%%%%%%%%%%%%%%%%%%%%%%%%%%%%%%%%%%%%%%%%%%%

\bibliographystyle{ACM-Reference-Format}
\bibliography{refs_CP}

%%%%%%%%%%%%%%%%%%%%%%%%%%%%%%%%%%%%%%%%%%%%%%%%%%%%%%%%%%%%%%%%%%%%%
%%%%%%%%%%%%%%%%%%%%%%%%%%%%%%%%%%%%%%%%%%%%%%%%%%%%%%%%%%%%%%%%%%%%%

\appendix

%%%%%%%%%%%%%%%%%%%%%%%%%%%%%%%%%%%%%%%%%%%%%%%%%%%%%%%%%%%%%%%%%%%%%%%%%%%%%%%
\section{Modules Details}
\label{ap:modules_details}
%%%%%%%%%%%%%%%%%%%%%%%%%%%%%%%%%%%%%%%%%%%%%%%%%%%%%%%%%%%%%%%%%%%%%%%%%%%%%%%

The following provides details about the initial set of modules
created for the Captive Portal education project.

\onecolumn

%%%%%%%%%%%%%%%%%%%%%%%%%%%%%%%%%%%%%%%%%%%%%%%%%%%%%%%%%%%%%%%%%%%%%%%%%%%%%%%

%% MODULE 1
\begin{table}[h!]
	\begin{tabular}{| p{0.3\textwidth} | p{0.3\textwidth} | p{0.3\textwidth} |}
	\hline
		\rowcolor{gray!20}\multicolumn{3}{l}{\bf Module 1: TCP Server and Client}
		\\
	\hline
	%\multicolumn{3}{l}{\textit{Overview}}	\\
	\multicolumn{3}{p{0.9\textwidth}}{This module lays the foundational understanding of TCP connections, vital for implementing
captive portal projects. It aims to solidify students’ network programming skills and introduce
them to Python coding for network applications.}
	\\
	\multicolumn{3}{p{0.9\textwidth}}{
	\textit{Objectives}
		\begin{itemize}
		\item Comprehend the architecture and function of TCP servers and clients.
		\item Understand the reason for using TCP to exchange information in Captive Portal.
		\item Have a general understanding of the entire Captive Portal framework.
		\end{itemize}
	}	\\
	\hline
	Pre-requisites	&	Implementations	& Learning Outcomes	\\
	\hline
	Basic understanding of TCP packets and python scripting.
	&
	Set up a TCP client in Mininet and enable it to communicate with a TCP server hosted in the virtual
machine.
	&
	Students will gain hands-on experience with network programming, enhancing their understanding
of how data is reliably transmitted across a network.
	\\
	\hline
\end{tabular}
\end{table}

%%%%%%%%%%%%%%%%%%%%%%%%%%%%%%%%%%%%%%%%%%%%%%%%%%%%%%%%%%%%%%%%%%%%%%%%%%%%%%%

%% MODULE 2
\begin{table}[h!]
	\begin{tabular}{| p{0.3\textwidth} | p{0.3\textwidth} | p{0.3\textwidth} |}
	\hline
		\rowcolor{gray!20}\multicolumn{3}{l}{\bf Module 2: SDN Switch}
		\\
	\hline
	%\multicolumn{3}{l}{\textit{Overview}}	\\
	\multicolumn{3}{p{0.9\textwidth}}{
Focuses on understanding Software-Defined Networking (SDN) controllers and the manipulation
of network traffic. Through hands-on practice with the POX SDN Controller, students will learn
how to program switches and process network packets.}
	\\
	\multicolumn{3}{p{0.9\textwidth}}{
		\textit{Objectives}
		\begin{itemize}
		\item Achieve proficiency in utilizing the POX controller to program a network switch.
		\item Explore the feasibility of modifying a standard switch to act as a captive portal.
		\end{itemize}
	}
	\\
	\hline
	Pre-requisites	&	Implementations	& Learning Outcomes	\\
	\hline
	Basic knowledge of networking concepts, including switches, MAC addresses, and packet flow.
	&
	Students are provided with a basic POX code framework and tasked with implementing a learning
switch using POX, thus applying their theoretical knowledge in a practical setting.
	&
	Enhanced understanding of SDN controllers and the ability to manipulate network traffic for
specific applications such as captive portals.
	\\
	\hline
\end{tabular}
\end{table}

%%%%%%%%%%%%%%%%%%%%%%%%%%%%%%%%%%%%%%%%%%%%%%%%%%%%%%%%%%%%%%%%%%%%%%%%%%%%%%%

%% MODULE 3
\begin{table}[h!]
	\begin{tabular}{| p{0.3\textwidth} | p{0.3\textwidth} | p{0.3\textwidth} |}
	\hline
		\rowcolor{gray!20}\multicolumn{3}{l}{\bf Module 3: DNS Server}
		\\
	\hline
	%\multicolumn{3}{l}{\textit{Overview}}	\\
	\multicolumn{3}{p{0.9\textwidth}}{
Introduces the functionality and implementation of DNS servers within the context of captive
portal technology. This module covers different approaches to handling DNS requests, emphasizing
hands-on experience with DNS manipulation.}
	\\
	\multicolumn{3}{p{0.9\textwidth}}{
		\textit{Objectives}
		\begin{itemize}
		\item Learn to uniformly respond to all DNS queries with a specific IP address.
		\item Develop skills to parse DNS requests, make requests on behalf of the user, and return genuine responses.
		\item Modify DNS request destination IPs using iptables for direct server responses.
		\end{itemize}
	}
	\\
	\hline
	Pre-requisites	&	Implementations	& Learning Outcomes	\\
	\hline
	Basic knowledge of UDP packets and DNS protocols.
	&
	Students can explore one of three different DNS configuration methods, from simple IP redirection
to complex request handling and response modification.
	&
	Students will understand how DNS servers can be manipulated to redirect or handle requests in a
captive portal setup.
	\\
	\hline
\end{tabular}
\end{table}

%%%%%%%%%%%%%%%%%%%%%%%%%%%%%%%%%%%%%%%%%%%%%%%%%%%%%%%%%%%%%%%%%%%%%%%%%%%%%%%

%% MODULE 4
\begin{table}[h!]
	\begin{tabular}{| p{0.3\textwidth} | p{0.3\textwidth} | p{0.3\textwidth} |}
	\hline
		\rowcolor{gray!20}\multicolumn{3}{l}{\bf Module 4: Web Server}
		\\
	\hline
	%\multicolumn{3}{l}{\textit{Overview}}	\\
	\multicolumn{3}{p{0.9\textwidth}}{
	Explores the deployment and configuration of a local web server to interact directly with network requests, crucial for the operation of a captive portal.
	}

	\\
	\multicolumn{3}{p{0.9\textwidth}}{
		\textit{Objectives}
		\begin{itemize}
			\item Set up and configure a local web server.
			\item Use iptables to redirect web requests to the local server.
		\end{itemize}
	}
	\\
	\hline
	Pre-requisites	&	Implementations	& Learning Outcomes	\\
	\hline
	Basic understanding of web server operation and network request handling.
	&
	Students will configure iptables to redirect web requests to a local
server, ensuring direct response capability.
	&
	Practical skills in web server setup and configuration, with an emphasis
on interaction with network requests in a captive portal context.
	\\
	\hline
\end{tabular}
\end{table}

%%%%%%%%%%%%%%%%%%%%%%%%%%%%%%%%%%%%%%%%%%%%%%%%%%%%%%%%%%%%%%%%%%%%%%%%%%%%%%%

%% MODULE 5
\begin{table}[h!]
	\begin{tabular}{| p{0.3\textwidth} | p{0.3\textwidth} | p{0.3\textwidth} |}
	\hline
		\rowcolor{gray!20}\multicolumn{3}{l}{\bf Module 5: Mininet Script}
		\\
	\hline
	%\multicolumn{3}{l}{\textit{Overview}}	\\
	\multicolumn{3}{p{0.9\textwidth}}{
	Delves into network simulation using Mininet, teaching students to
construct and modify network topologies to route user requests through
specific servers, mimicking a captive portal environment.
	}
	\\
	\multicolumn{3}{p{0.9\textwidth}}{
		\textit{Objectives}
		\begin{itemize}
			\item Understand network structure and the necessary modifications for captive portal functionality.
			\item Learn to use Mininet for creating and testing network topologies.
		\end{itemize}
	}
	\\
	\hline
	Pre-requisites	&	Implementations	& Learning Outcomes	\\
	\hline
	Basic knowledge of network topology and mininet.
	&
	Students have the flexibility to design the network according to the
	requirements of the captive portal setup they envision.
	&
	Ability to simulate and test network configurations, crucial for the
	development and deployment of captive portal technologies.
	\\
	\hline
\end{tabular}
\end{table}

%%%%%%%%%%%%%%%%%%%%%%%%%%%%%%%%%%%%%%%%%%%%%%%%%%%%%%%%%%%%%%%%%%%%%%%%%%%%%%%

\twocolumn

%%%%%%%%%%%%%%%%%%%%%%%%%%%%%%%%%%%%%%%%%%%%%%%%%%%%%%%%%%%%%%%%%%%%%%%%%%%%%%%
%%%%%%%%%%%%%%%%%%%%%%%%%%%%%%%%%%%%%%%%%%%%%%%%%%%%%%%%%%%%%%%%%%%%%%%%%%%%%%%

%%%%%%%%%%%%%%%%%%%%%%%%%%%%%%%%%%%%%%%%%%%%%%%%%%%%%%%%%%%%%%%%%%%%%

\section{Appendix: Manual configuration \& GUI attachment}

%%%%%%%%%%%%%%%%%%%%%%%%%%%%%%%%%%%%%%%%%%%%%%%%%%%%%%%%%%%%%%%%%%%%%

%For advanced courses and users of mininet, if they prefer to gain
%deeper insights of the technology and procedural setup, instead of using
%our provided configuration and setup scripts,
%they could follow the steps described in Fig.~\ref{fig:mininet_CP_demo}.
%
%This would require the usual mininet configuration and setup steps:

\begin{figure}
	\begin{subfigure}{\columnwidth}
		\includegraphics[width=\columnwidth]{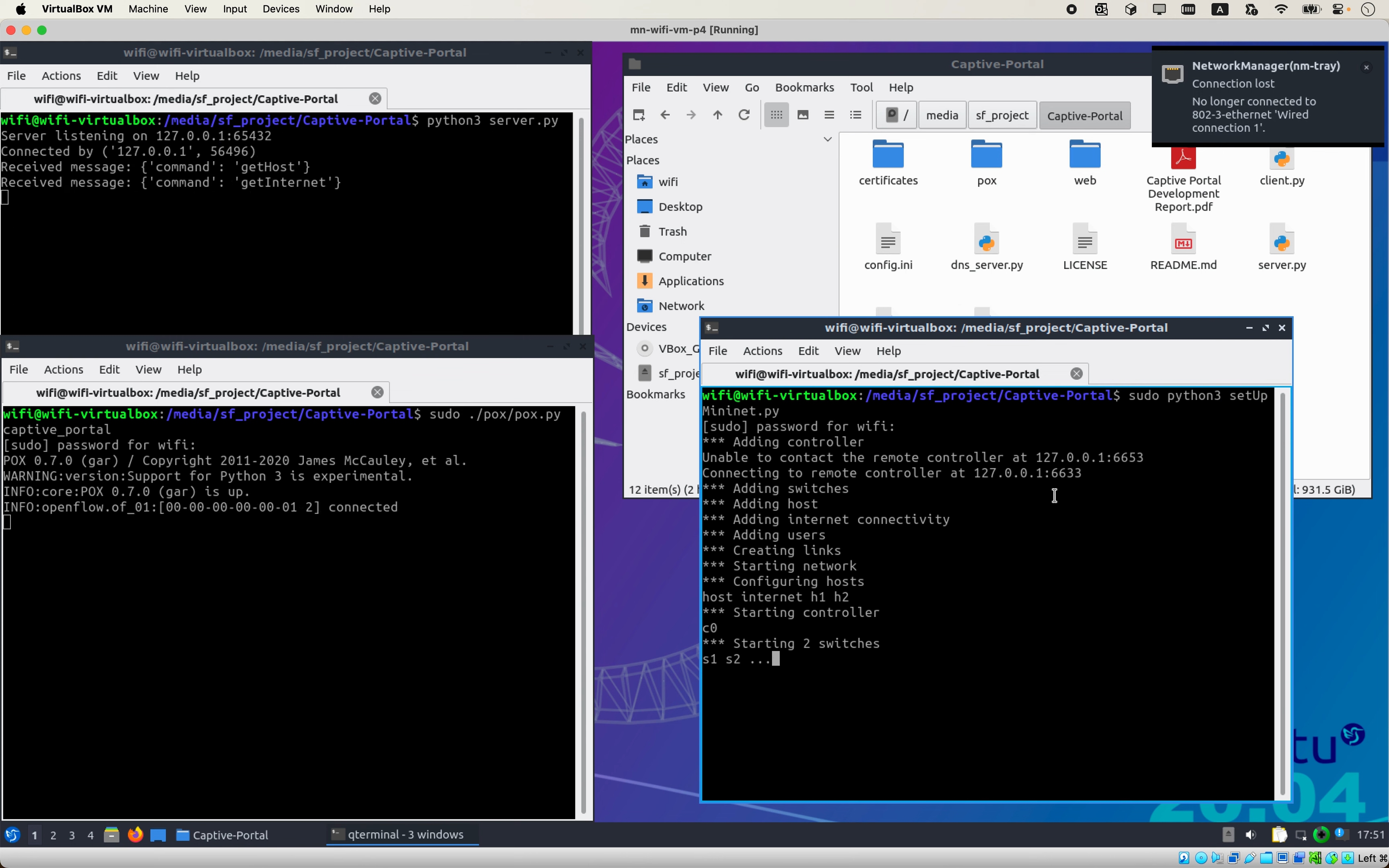}
		\caption{Configuration setup in a VM for running mininet as a Captive Portal.}
		\label{fig:mininet_CP_setup}
	\end{subfigure}
	\begin{subfigure}{\columnwidth}
		\includegraphics[width=\columnwidth]{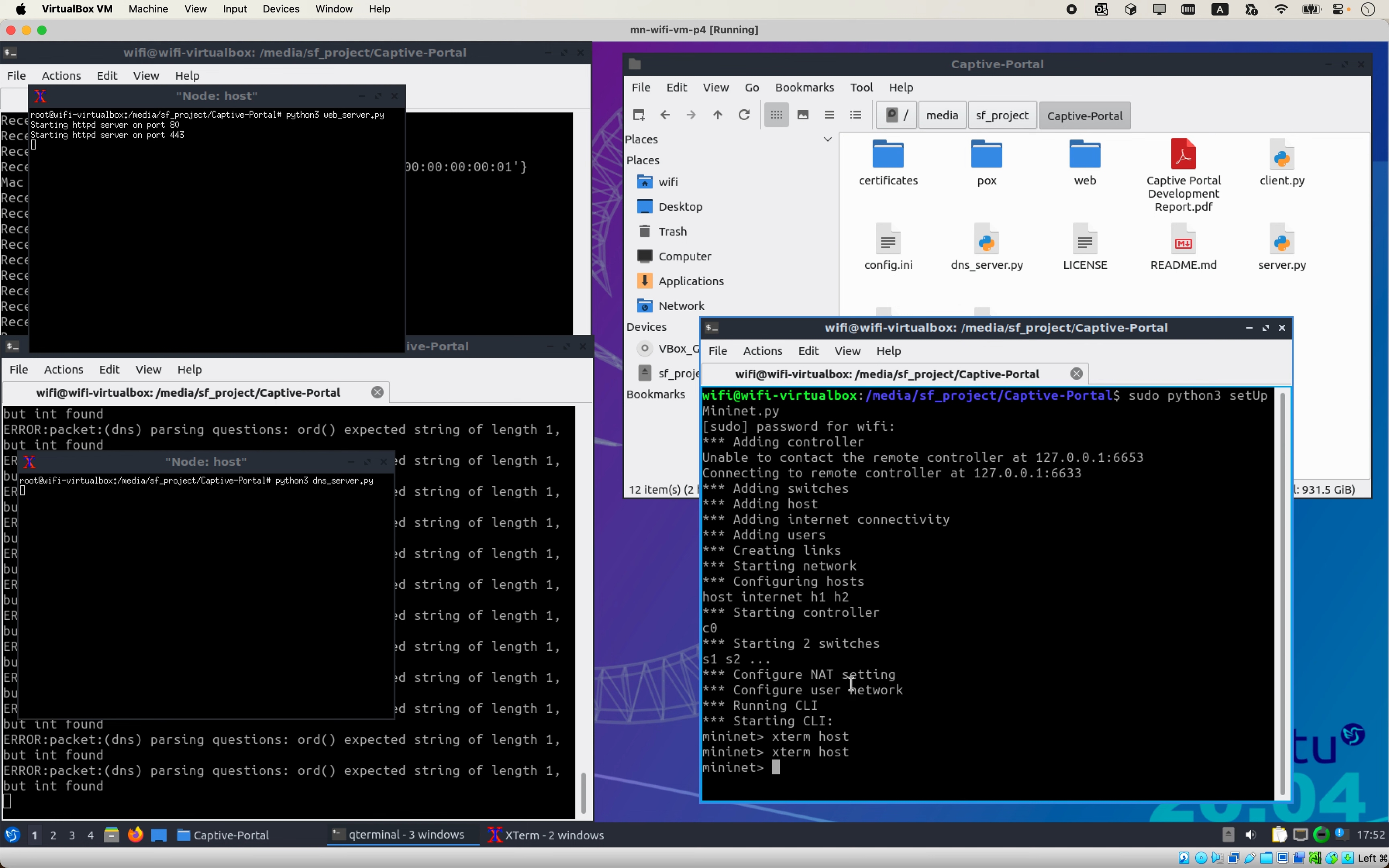}
		\caption{Setup of DNS server with the mininet Captive Portal configuration.}
		\label{fig:mininet_CP_dnsserver}
	\end{subfigure}
%\end{figure}
%\begin{figure}[ht]\ContinuedFloat
	\begin{subfigure}{\columnwidth}
		\includegraphics[width=\columnwidth]{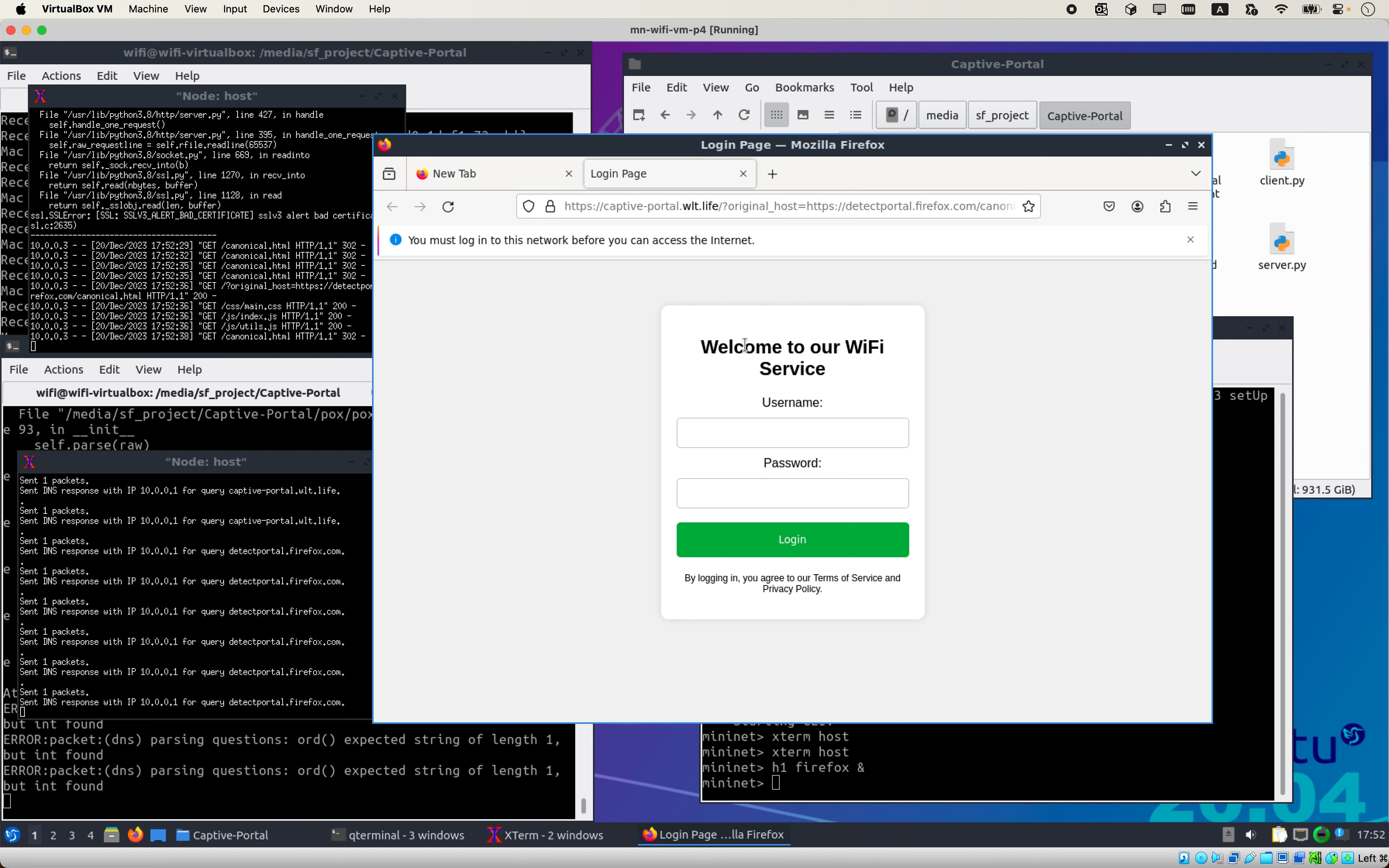}
		\caption{Demonstration of the DNS server interception of network traffic via a web-browser request.}
		\label{fig:mininet_CP_firefox}
	\end{subfigure}

	\caption{Demonstration of the sequence of steps required to setup and execute the DNS spoofing technology to forge the Captive Portal allowance.}
	\label{fig:mininet_CP_demo}
\end{figure}

For advanced courses and users of Mininet who wish to gain deeper insights into
the technology and procedural setup, the following steps provide a brief
introduction to run Mininet with a graphical user interface and additional
networking tools:

\begin{enumerate}
	\item Initiate the virtual environment that hosts the Mininet installation.
	\item Start the POX controller and server setup as shown in Fig.~\ref{fig:mininet_CP_setup}.
	\item Open additional terminals or sessions
		to run the corresponding web and DNS servers as in Fig.~\ref{fig:mininet_CP_dnsserver}.
	\item Launch an X-terminal to run a web-browser and optionally a network packet 
		sniffer, such as \texttt{wireshark}, as illustratged in Fig.~\ref{fig:mininet_CP_firefox}.
\end{enumerate}

%This would also allow to use any other mininet base-configuration -- for instance,
%including an X-server which would allow them to run GUIs and trigger the 
%typical pop-ups from browsers or even capturing network traffic using tools such as
%\texttt{wireshark}\footnote{\url{https://www.wireshark.org/}}, or even using a
%wireless mininet setup \cite{10.1145/2934872.2959070}.

This process allows users to manually configure and customize their Mininet
setup, including the option to run graphical user interfaces (GUIs), trigger
browser pop-ups typical of Captive Portals, or capture network traffic using
\texttt{Wireshark}\footnote{\url{https://www.wireshark.org/}}.
It also enables users to explore other Mininet base configurations,
such as wireless mininet setup \cite{10.1145/2934872.2959070}.

However, it is important to note that these steps are meant to demonstrate what
the final product will look like once students have completed all or part of
their project code, regardless of how many modules the students complete, the
final product will be a fully functional Captive Portal, it is not an exact
reflection of how students will test their code.

Notice that to properly demonstrate these capabilities, some additional configuration
steps are required beyond what is shown in the images. These additional steps,
as well as more specific configuration details, will be provided in the
project's GitHub repository.

%%%%%%%%%%%%%%%%%%%%%%%%%%%%%%%%%%%%%%%%%%%%%%%%%%%%%%%%%%%%%%%%%%%%%
%%%%%%%%%%%%%%%%%%%%%%%%%%%%%%%%%%%%%%%%%%%%%%%%%%%%%%%%%%%%%%%%%%%%%

%%%%%%%%%%%%%%%%%%%%%%%%%%%%%%%%%%%%%%%%%%%%%%%%%%%%%%%%%%%%%%%%%%%%%
%%%%%%%%%%%%%%%%%%%%%%%%%%%%%%%%%%%%%%%%%%%%%%%%%%%%%%%%%%%%%%%%%%%%%

\end{document}